\documentclass[twocolumn,showpacs,preprintnumbers,amsmath,amssymb,prl,aps,superscriptaddress]{revtex4-1}
\usepackage{graphicx}
\usepackage{amsmath}
\usepackage{verbatim}

\newcommand{\ket}[1]{\left\lvert #1 \right\rangle}
\newcommand{\EJ}{E_J}
\newcommand{\EC}{E_C}
\newcommand{\s}{\mathrm{s}}
\newcommand{\ms}{\mathrm{ms}}
\newcommand{\us}{\mu\mathrm{s}}
\newcommand{\ns}{\mathrm{ns}}
\newcommand{\ueV}{\mu\mathrm{eV}}

\newcommand{\DeltaAl}{\Delta}

\newcommand{\Rppkk}{R_{kk'}(\tau)}
\newcommand{\Rpg}{R_{00}(\tau)}
\newcommand{\Rpe}{R_{11}(\tau)}
\newcommand{\Rpeg}{R_{10}(\tau)}
\newcommand{\Rpegz}{R_{10}(\tau\to 0)}
\newcommand{\wzo}{\omega_{01}}
\newcommand{\kHz}{\mathrm{kHz}}
\newcommand{\MHz}{\mathrm{MHz}}
\newcommand{\GHz}{\mathrm{GHz}}
\newcommand{\Tf}{T_r}
\newcommand{\mK}{\mathrm{mK}}
\newcommand{\df}{\Delta f}
\newcommand{\dT}{\tau}
\newcommand{\dt}{\Delta t}
\newcommand{\Ma}{M_0}
\newcommand{\Mb}{M_1}
\newcommand{\Mc}{M_2}
\newcommand{\Md}{M_3}
\newcommand{\Gextot}{\Gamma_{01}}
\newcommand{\Goee}{\Gamma^{eo}_{11}}
\newcommand{\Goeg}{\Gamma^{eo}_{00}}
\newcommand{\Goer}{\Gamma^{eo}_{10}}
\newcommand{\Goeex}{\Gamma^{eo}_{01}}
\newcommand{\Gex}{\Gamma^{ee}_{01}}
\newcommand{\Gr}{\Gamma^{ee}_{10}}
\newcommand{\Grt}{\Gamma_\mathrm{rts}}
\newcommand{\Grtot}{\Gamma_{10}}
\newcommand{\um}{\mu \mathrm{m}}
\newcommand{\fev}{f_e}
\newcommand{\fod}{f_o}
\newcommand{\K}{\mathrm{K}}
\newcommand{\kohm}{\mathrm{k}\Omega}
\newcommand{\wrr}{\omega_r}
\newcommand{\Ttwostar}{T_{2}^{\ast}}

\begin{document}

\title{Millisecond charge-parity fluctuations and induced decoherence in a superconducting qubit}

\author{D.~Rist\`e}
\author{C.~C.~Bultink}
\author{M.~J.~Tiggelman}
\author{R.~N.~Schouten}
\affiliation{Kavli Institute of Nanoscience, Delft University of Technology, P.O. Box 5046, 2600 GA Delft, The Netherlands}
\author {K.~W.~Lehnert}
\affiliation{JILA, National Institute of Standards and Technology and Department of Physics, University of Colorado, Boulder, Colorado 80309, USA}
\author{L.~DiCarlo}
\affiliation{Kavli Institute of Nanoscience, Delft University of Technology, P.O. Box 5046,
2600 GA Delft, The Netherlands}
\date{\today}

\begin{abstract}
Quasiparticle excitations adversely affect the performance of superconducting devices in a wide range of applications. 
They limit the sensitivity of photon detectors in astronomy~\cite{Day03,Stone12}, the accuracy of current sources in metrology~\cite{Pekola12}, the cooling power of micro-refrigerators~\cite{Giazotto06}, and could break the topological protection of Majorana qubits~\cite{Lutchyn10}. 
In superconducting circuits for quantum information processing, tunneling of quasiparticles across Josephson junctions constitutes a decoherence mechanism~\cite{Martinis09,Lenander11,Catelani11,Catelani12}. As relaxation and pure dephasing times of transmon-type charge qubits now reach $100~\us$, understanding whether quasiparticle tunneling may already bottleneck coherence is of high interest. 
We integrate recent advances in qubit readout~\cite{Riste12} and feedback control~\cite{Riste12b} in circuit quantum electrodynamics~\cite{Wallraff04} to perform the first real-time observation of quasiparticle tunneling in a transmon qubit. We demonstrate  quasiparticle-tunneling contributions to qubit relaxation and pure dephasing in the millisecond range. Thus, quasiparticle tunneling will not limit coherence for at least  one order of magnitude beyond the state of the art. 
\end{abstract}

\maketitle

The preservation of  charge parity (even or odd number of electrons) has historically been of primary concern in superconducting quantum information processing (QIP). In the first superconducting qubit, termed Cooper pair box (CPB)~\cite{Bouchiat98}, maintaining the parity in a small island connected to a reservoir via Josephson junctions is essential to qubit operation. The qubit states $\ket{0}$ and $\ket{1}$ consist of symmetric superpositions of charge states of equal parity,  brought into resonance by a  controlled charge bias $n_g$ and split by the Josephson tunneling energy $\EJ$ ($\lesssim \EC$, the island Cooper-pair charging energy). Quasiparticle (QP) tunneling across the junction changes the island parity, ``poisoning'' the box until  parity switches back or $n_g$ is offset by $\pm e$~\cite{Lutchyn06}. QP poisoning in CPBs has been extensively studied, with most experiments~\cite{Ferguson06,Naaman06,Court08,Shaw08} finding  parity switching times of $10~\us - 1~\ms$, and some $>1~\s$~\cite{Tuominen92,Eiles94,Amar94}.  While these times are long compared to qubit gate operations ($\sim10~\ns$), the sensitivity of the CPB qubit transition frequency $\wzo$ to background charge fluctuations  limits the  dephasing time to $<1~\us$, severely restricting the use of traditional CPBs in QIP. 

Engineering the CPB into the transmon regime $\EJ\gg\EC$~\cite{Koch07,Schreier08} exponentially suppresses the sensitivity of $\wzo$ to charge-parity and background charge fluctuations. However, recent theory~\cite{Martinis09,Catelani11,Catelani12} predicts that QP tunneling remains a relevant source of qubit relaxation and pure dephasing, particularly as  improved understanding of dielectric loss~\cite{Paik11} and the Purcell effect~\cite{Houck08} has allowed reaching the $100~\us$ scale in cQED. To guide further improvements, it is imperative to precisely pinpoint the timescale for QP tunneling and its contribution to qubit decoherence. To date, only upper and lower bounds have been placed~\cite{Schreier08,Sun12}.  

\begin{figure*}
\includegraphics[width=1.42\columnwidth]{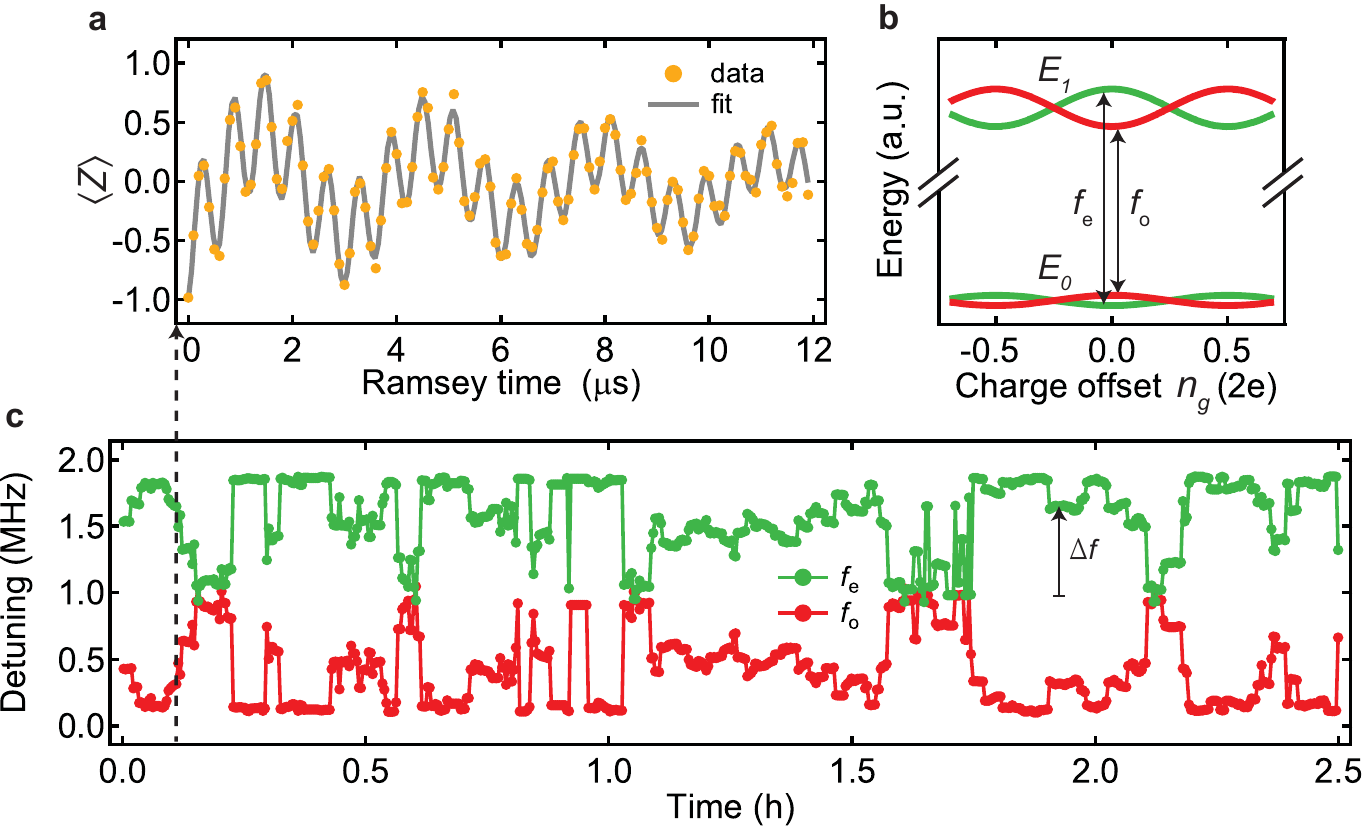}
\caption{{\bf Bistability and drift of the qubit transition frequency.} {\bf a,} Ramsey fringe experiment (dots) and  best-fit sum of two decaying sinusoids (curve). The reference oscillator is detuned $1~\MHz$ from the average qubit transition frequency $\wzo/2\pi=4.387~\GHz$. {\bf b,} Sketch of the charge dispersion of the first two levels of the transmon qubit, showing $2e$ periodicity. 
QP tunneling across the junction shifts $n_g$ precisely by $e$, resulting in two transition frequencies  $\fev$ and $\fod$ (not to scale). {\bf c}, 
Repeated Ramsey experiments ($15~\s$ each) show a symmetric drift of $\fev$ and $\fod$ around $\wzo/2\pi$, arising from background charge motion.
The frequency difference $2\df=\fev-\fod$ ranges from $0$ to $1.76~\MHz$ (see also the Supplementary Methods).}
\end{figure*}

Here, we transform a state-of-the-art  single-junction  transmon qubit into a real-time charge-parity detector, and demonstrate that QP tunneling does not preclude reaching the millisecond timescale in transmon qubit coherence. Our qubit is controlled and measured in a 3D circuit quantum electrodynamics (cQED)  architecture~\cite{Paik11}, i.e., the same  environment that the qubit experiences in QIP applications, without need for any additional electrometer or other circuitry. At the heart of our detection scheme is a very small but detectable parity dependence of the qubit transition frequency (up to $0.04\%$ of the average $\wzo/2\pi = 4.387~\GHz$), obtained by choosing $\EJ/\EC=25$.

Standard Ramsey fringe experiments provide the first evidence of QP tunneling across the qubit junction, as shown in Fig.~1 for a refrigerator temperature $\Tf=20~\mK$. 
Instead of the usual decaying sinusoid, we observe two. Repeated Ramsey experiments always reveal two frequencies, fluctuating symmetrically about the average $\wzo$ (Fig.~1c). The double frequency pattern results from the qubit frequency sensitivity to charge, with QP tunneling events shifting the energy levels by $\pm e$ in the $n_g$ axis. The fluctuation in the frequency difference $\df$ is due to background charge motion slow compared to QP tunneling. 
The observation of two frequencies in every experiment shows that QP tunneling is fast compared to the averaging time ($\sim\!15~\s$), but slow compared to the maximum $1/2\df\sim 5~\us$~\cite{Paladino02}. 
From the similar amplitude of the sinusoids, we infer that the two parities are equally likely. 
Clearly, these time-averaged measurements only loosely bound the timescale for QP tunneling, similarly to Refs.~\onlinecite{Schreier08,Sun12}.

In order to accurately pinpoint the timescale for QP tunneling, we have devised a scheme to monitor the charge parity in real time (Fig.~2a), taking advantage of recent developments in high-fidelity nondemolition readout~\cite{Riste12} and feedback control~\cite{Riste12b}. 
Starting from $\ket{0}$, the qubit is prepared in the superposition state ($\ket{0}+\ket{1})/\sqrt{2}$ with a $\pi/2$ $y$ pulse at $\wzo$. 
The qubit then acquires a phase $\pm\pi/2$  during a chosen idle time $\dt=1/4\df$, where the $+$ $(-)$ sign corresponds to even (odd) parity.
A second $\pi/2$ $x$ pulse completes the mapping of parity into a qubit basis state, even $\to \ket{0}$, odd $\to \ket{1}$. A following projective qubit  measurement~\cite{Riste12} ideally matches the result $M=1$ $(-1)$ to even (odd) parity. Feedback-based reset~\cite{Riste12b} reinitializes the qubit to $\ket{0}$ and allows repeating this sequence every $\dt_\mathrm{exp}=6~\us$. We note that this scheme realizes the charge-parity meter in the envisioned top-transmon architecture~\cite{Hassler11}, in which a transmon is used to manipulate and readout Majorana qubits. 

\begin{figure*}
\includegraphics[width=2\columnwidth]{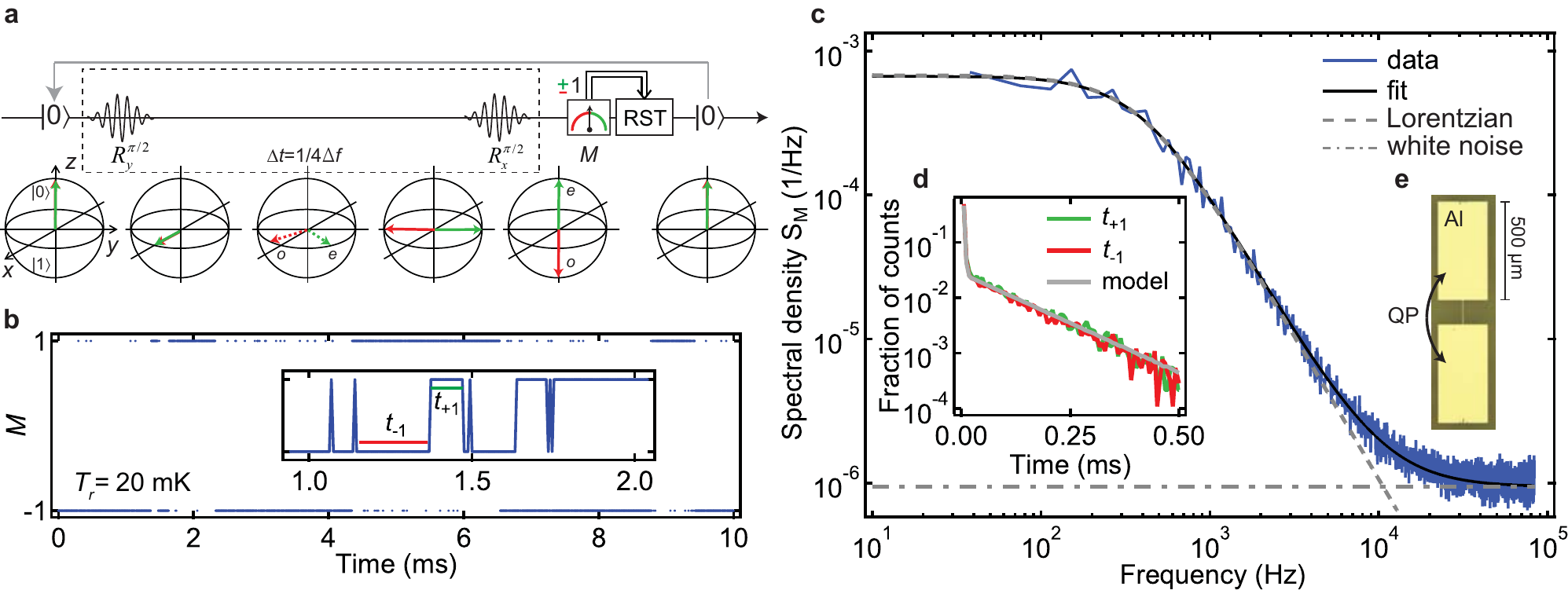}
\caption{{\bf Real-time measurement of QP tunneling.} {\bf a,} Ramsey-type sequence converting the qubit into a charge-parity detector.
The sequence (see main text) is equivalent to a $\pi$ rotation conditioned on odd parity. 
{\bf b,} Snapshots of a typical measurement trace, $48~\ms$ long. Dots are repetitions of the experiment in {\bf a}, at $6~\us$ interval. {\bf c,}  Double-sided power spectral density of $M$, obtained by averaging the squared Fourier transform of 45 consecutive repetitions of {\bf b}. The best fit of equation~\eqref{eq:lor} gives $1/\Grt=0.794\pm 0.005~\ms$. Repeated experiments have a standard deviation of $0.09~\ms$. {\bf d,} Histograms of dwell times for $M=\pm1$. The grey curve is a model of RTS with symmetric rate $\Grt$ and detection fidelity $F$, extracted from the fit in the main panel. {\bf e}, Optical image of a qubit with identical geometry~\cite{Riste12} to that used in this experiment.  See Supplementary Methods for several control experiments testing the measurement protocol.}
\end{figure*}

The time evolution of charge parity is encoded in the series of results $M$ (Fig~2b). 
The time series has zero average, confirming that the two charge parities are equally probable. 
Both the QP dynamics and the detection infidelity determine the distribution of  dwell times $t_1$ and $t_{-1}$ (Fig.~2d).  The measured identical histograms match a numerical simulation of a symmetric random telegraph signal (RTS) with transition rate $\Grt$, masked by uncorrelated detection errors occurring with probability $(1-F)/2$.  These two noise processes contribute distinct signatures to the spectral density of $M$ (Fig.~2c). The best fit of the form
\begin{equation}
\label{eq:lor} 
S_M(f) = F^2 \frac{4\Grt}{(2\Grt)^2+(2\pi f)^2} + (1-F^2) \Delta t_{\mathrm{exp}}
\end{equation}
shows excellent agreement, giving $1/\Grt=0.79~\ms$  and $F = 0.92$. 

\begin{figure} 
\includegraphics[width=\columnwidth]{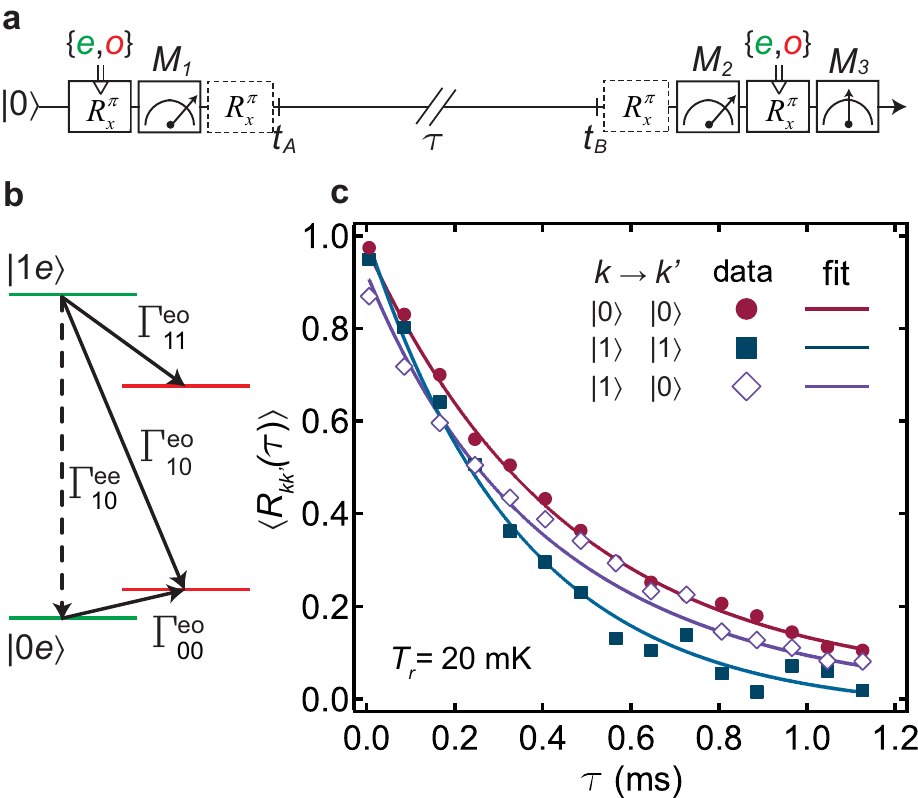}
\caption{{\bf Rates of QP tunneling with and without qubit transitions.} {\bf a,} Pulse sequence measuring the autocorrelation function of charge parity. Two parity measurements $\Mb$ and $\Md$ (see Fig.~2) are separated by a waiting time $\dT$. Postselection on $\Mb=1$~\cite{Riste12} prepares the state $\ket{0e}$. Similarly, a measurement $\Mc=1$ at the end of $\dT$ ensures that the final qubit state is $\ket{0}$. $\Md$ will coincide with $\Mb$ only if the parity is unchanged. Inserting $\pi$ rotations after $\Mb$ and/or before $\Mc$ allows measuring the parity autocorrelation for different combinations of qubit states. A preliminary measurement $\Ma$ (not shown) initializes the qubit in $\ket{0}$ by postselection. {\bf b,} Diagram of the four energy levels with the modeled transition rates  (not to scale). {\bf c,} Charge-parity autocorrelation $\Rppkk$ for qubit in state $\ket{0}$ (dots), $\ket{1}$ (squares), or having relaxed from $\ket{1}$ to $\ket{0}$ (diamonds) during $\dT$. 
The average of the conditioned $\Md$ is corrected for detection  infidelity (see Methods). 
Fitting the solution of the rate equations, conditioned on initial and final qubit state, gives the inverse rates: $1/\Gamma_{00}^{eo}=0.92\pm0.04~\ms$, $1/\Gamma_{11}^{eo}=0.70\pm0.06~\ms$, $1/\Gamma^{ee}_{10}=0.14\pm0.06~\ms$, $1/\Gamma_{10}^{eo}=3.3\pm1.0~\ms$.} 
\end{figure}

While the above scheme detects a characteristic time for QP tunneling, additional experiments are needed to distinguish QP tunneling events that cause qubit transitions from those that do not. 
For this purpose, we model the system with four levels $\ket{kl}$ ($k$ and $l$ denote the qubit and parity state, respectively), and rates  $\Gamma_{kk'}^{ll'}$ connecting them, with $k$ $(k')$ and $l$ $(l')$ the initial (final) qubit and parity state, respectively (Fig.~3b). Based on the identical distributions of dwell times, we consider symmetric rates $\Gamma_{kk'}^{eo} = \Gamma_{kk'}^{oe}$. 

To extract the rates, we measure the autocorrelation function of charge parity, conditioned on specific initial and final qubit states (Fig.~3). 
Conditioning on a first charge-parity measurement $\Mb=+1$ postselects the qubit in $\ket{0}$ and even parity. 
A second measurement $\Mc$ follows a waiting time $\dT$.
By conditioning also on $\Mc=+1$, we ensure that the qubit both starts and ends in $\ket{0}$. A second parity measurement, ending with $\Md$, completes the sequence. The average result, once corrected for detector infidelity (see Methods), is the parity autocorrelation $\Rpg = \langle P(0)P(\dT) \rangle_{00}$, with first (second) subscript indicating initial (final) qubit state.  Neglecting qubit excitation, i.e., setting $\Gextot=\Gex+\Goeex=0$, $\Rpg$ simply decays as $\exp(-2\Goeg t)$. The exact solution  shows that this remains a valid approximation when including the measured $\Gextot = 1/6~\ms^{-1}$, since the probability of multiple qubit transitions in $\dT$ is negligible.
Similarly, we measure the parity autocorrelation with qubit initially and finally in $\ket{1}$,  $\Rpe\approx\exp(-2\Goee t)$. 
To do this, we use the same conditioning, but apply a $\pi$ pulse after $\Mb$ and before $\Mc$. Exponential decay fits give $1/\Goeg = 0.92\pm0.04~\ms$ and $1/\Goee = 0.70\pm0.06~\ms$.  

To quantify the contribution of QP tunneling to the measured qubit relaxation time $T_1=1/\Grtot=0.14~\ms$, we apply the same method, but condition on initial state $\ket{1}$ and final state $\ket{0}$.  The ratio of QP-induced to total relaxation rates $\alpha\equiv \Goer/\Grtot$ ($\Grtot=\Gr+\Goer$) can be extracted from $\Rpegz = 1-2\alpha$. 
The best fit of the model $\Rpeg$ to the data, with $\alpha$  as only free  parameter, 
gives $1/\Goer = 3.3\pm1.0~\ms$ and $1/\Gr = 0.14\pm0.06~\ms$.
This result clearly demonstrates that QP tunneling does not dominate qubit relaxation at $\Tf = 20~\mK$, contributing only $5\%$ of qubit relaxation events. 

\begin{figure}
\includegraphics[width=\columnwidth]{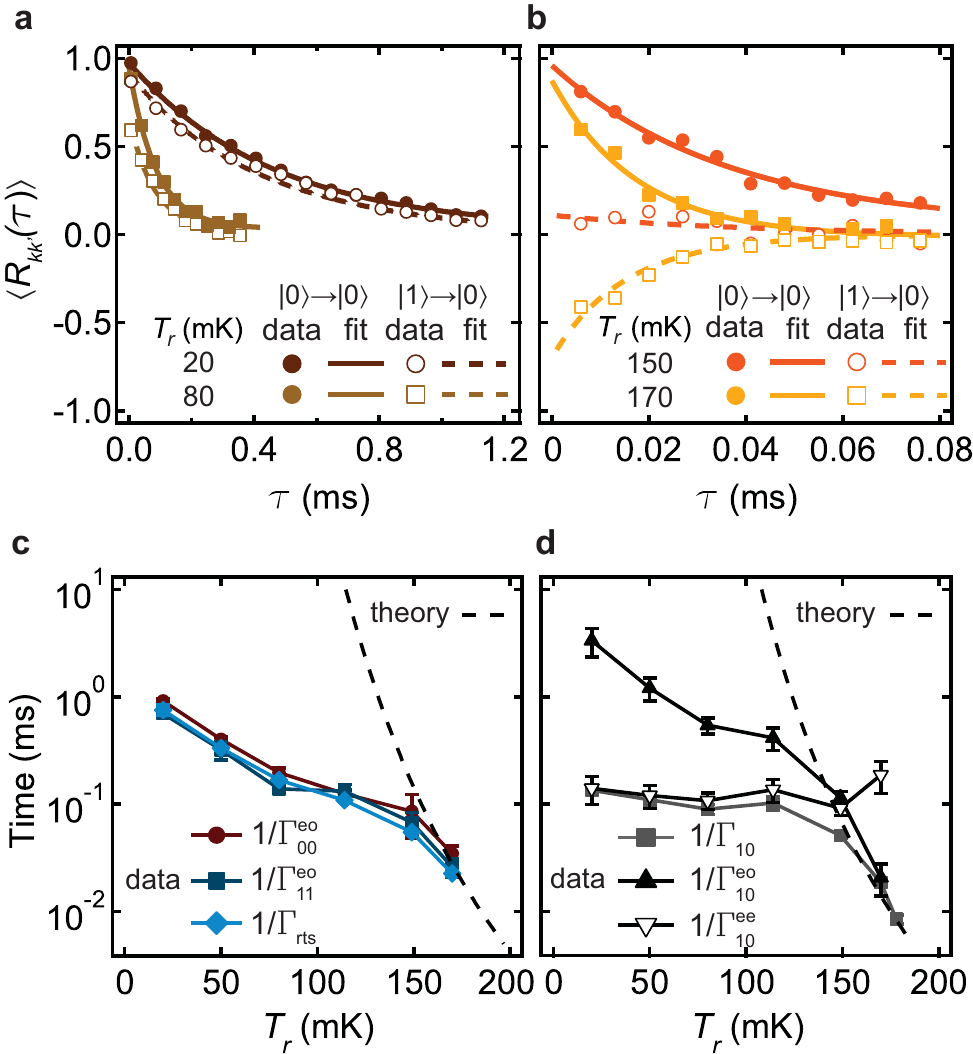}
\caption{{\bf Temperature dependence of QP tunneling times.} 
{\bf a,b,} Charge-parity autocorrelation functions 
$\Rpg$ and $\Rpeg$ at $20, 80$ ({\bf a}), $150$ and $170~\mK$ ({\bf b}). 
$\Rpegz$ progressively decreases, indicating an increasing contribution of QP tunneling to qubit relaxation. {\bf c,} QP tunneling times for the ground-state ($1/\Goeg$, dots) and excited-state ($1/\Goee$, squares) manifold extracted from $\Rpg$ and  $\Rpe$ (not shown). Fits to $S_M(f)$ (Supplementary Methods, Fig.~S5) give similar values for $1/\Grt$  (diamonds). Dashed curve: theory for $\Goeg$~\cite{Martinis09,Catelani12} for thermally distributed QPs and $\DeltaAl = 170~\ueV$. {\bf d,} Relaxation times with ($1/\Goer$, upward triangles) and without ($1/\Gr$, downward triangles) QP tunneling, obtained from $\Rpeg$ and the overall relaxation time $T_1=1/\Grtot$ (squares). Dashed curve: equation~\eqref{eq:rel_theo} for thermal equilibrium. Error bars are 1 s.d.} 
\end{figure}

To facilitate comparison to  theory, we perform the above experiments at elevated $\Tf$ (Fig.~4). We observe that $\Grt$, $\Goeg$, $\Goee$, and $\Goer$ have similar magnitude and jointly increase with $\Tf$ in the range $20-170~\mK$. 
However, $T_1$ remains insensitive to $\Tf$ until $150~\mK$. The observed sign reversal in $\Rpegz$ near this temperature  (Fig.~4b) indicates that QP tunneling becomes the dominant relaxation process. 

The effect of QP dynamics on the qubit degree of freedom in superconducting circuits has been extensively studied theoretically~\cite{Lutchyn06,Martinis09,Catelani11,Catelani12}. 
For transmon qubits, the predicted QP-induced relaxation rate is~\cite{Martinis09,Catelani11}
\begin{equation}
\label{eq:rel_theo}
\Goer \approx \frac{x_\mathrm{qp}}{\pi}\sqrt{2\DeltaAl\wzo},
\end{equation}
where $x_\mathrm{qp}=  n_\mathrm{qp}/2\nu_0\DeltaAl$ is the QP density $n_\mathrm{qp}$ normalized to the Cooper-pair density, with $\nu_0=1.2\times10^{4}~\um^{-3}\ueV^{-1}$ the single-spin density of states at the Fermi energy~\cite{Court08} and $\DeltaAl$ the Al superconducting gap. This relation holds for any energy distribution of QPs. 
For $\Tf\geq150~\mK$, the data closely match equation~\eqref{eq:rel_theo} using the thermal equilibrium $x_\mathrm{qp}  = \sqrt{2\pi\Tf/\DeltaAl}e^{-\DeltaAl/\Tf}$ and $\DeltaAl = 170~\ueV$, the value estimated from the normal-state resistance of the junction (see Methods). 
The suppression of $\Goer$ at lower $\Tf$ is much weaker than expected from a thermal QP distribution. 
Using equation~\eqref{eq:rel_theo}, we estimate $n_\mathrm{qp} = 0.04\pm0.01~\um^{-3}$ at $\Tf=20~\mK$, matching the lowest value reported for Al in a Cooper-pair transistor for use in metrology~\cite{Saira12}. Improved shielding against infrared radiation~\cite{Barends11} could further decrease $n_\mathrm{qp}$ at low $\Tf$, and will be pursued in future work. 

QP tunneling events that do not induce qubit transitions contribute to pure qubit dephasing.  Calculations based on  Refs.~\onlinecite{Martinis09} and~\onlinecite{Catelani12} predict $\Gamma^{eo}_{kk}\approx\Goer$, in good agreement with the data (Fig.~4c). It is presently not understood whether these QP tunneling events completely destroy qubit superposition states (case A)  or simply change the qubit precession frequency (case B).  In either case,  in the regime of strongly coupled RTS valid for our experiment  ($\Goeg,\Goee\ll \df$~\cite{Paladino02}) the QP-induced dephasing time is $2/(\Gamma^{eo}_{00}+\Gamma^{eo}_{11})\sim0.8~\ms$. For case B, this time would further increase in the weak-coupling regime (attained at $\EJ/\EC\gtrsim 60$) due to motional averaging~\cite{Paladino02}.

In conclusion, we have measured the characteristic times of QP tunneling across the single junction of a 3D transmon by converting the qubit into a real-time charge-parity detector. First, probing charge parity every $6~\us$ with a Ramsey-like sequence reveals a symmetric RTS with $0.8~\ms$ characteristic switching time. Second, measuring the charge-parity autocorrelation function, conditioned on specific initial and final qubit states, 
distinguishes QP tunneling that induces qubit relaxation from that which does not. 
We have shown that QP tunneling is not the dominant relaxation mechanism, contributing just $5\%$ of relaxation events in our state-of-the-art transmon. Reaching the millisecond horizon in coherence will facilitate the realization of fault-tolerant QIP with superconducting circuits. 

\section{methods}

{\it Device parameters.} The transmon has Josephson energy $\EJ = 8.442~\GHz$ and charging energy $\EC=0.334~\GHz$. Using the Ambegaokar-Baratoff relation $\EJ R_{n} = 
 \Delta/8e^2$ and the measured room-temperature resistance $R_{n,300\K}=15.2~\kohm$ of the single Josephson junction, we estimate $\Delta = 170~\ueV$. The qubit couples to the
fundamental mode of the cavity $\wrr/2\pi = 6.551~\GHz$ (decay rate $\kappa/2\pi = 720~\kHz$) with strength $g/2\pi = 66~\MHz$, inducing a dispersive shift $2\chi/2\pi=-1.0~\MHz$. The qubit relaxation time $T_1$ may be limited by the Purcell effect~\cite{Houck08}. A simple estimate including only the fundamental cavity mode gives $240~\us$. The dephasing time, $\Ttwostar = 10-25~\us$, is limited by background charge fluctuations (see Supplementary Methods).

{\it Experimental setup.}
Projective readout with $99\%$ fidelity is achieved by homodyne detection with a $400~\ns$ pulse at $\wrr-\chi$, aided by a Josephson parametric amplifier~\cite{Riste12}. The qubit reset is implemented with a home-built feedback controller based on a complex programmable logic device (Altera MAX V)  that integrates the last $200~\ns$ of the readout signal and conditionally triggers a $\pi$ pulse (Gaussian, $\sigma = 8~\ns$, $32~\ns$ long) $2~\us$ after the end of the measurement~\cite{Riste12b} (see the Supplementary Methods for more details). 

{\it Extraction of QP tunneling rates.} To convert $\langle \Md (\dT) \rangle_{kk'}$ into $\Rppkk$, we correct for the overall detection errors, distributed among readout ($<1\%$) and reset  ($\sim\!1\%$) infidelities, suboptimal $\dt$ ($<2\%$), and dephasing during $\dt$ (remaining $1-3\%$). For this correction, we first fit an exponential decay to $\langle \Md(\dT) \rangle_{00}$ and $\langle \Md(\dT) \rangle_{11}$. The average of the best-fit value at $\dT=0$ is used to renormalize the data in Figs.~3c  and 4a,b. The fitted decay times are $1/2\Goeg$ and $1/2\Goee$, respectively. To extract $\Goer$ and $\Gr$, we fit the solution of equation~(2) to $\Rpeg$, using $\Goer+\Gr=\Grtot$. 
$\Grtot$ is obtained from 
the equilibration time $T_\mathrm{eq}$ after inverting the steady-state populations  $P_{\ket{0},\mathrm{ss}}, P_{\ket{1},\mathrm{ss}}$ with a $\pi$ pulse: 
\begin{equation}
\label{eq:gamma01}
\Grtot = \frac{P_{\ket{0},\mathrm{ss}}}{(P_{\ket{0},\mathrm{ss}}+P_{\ket{1},\mathrm{ss}}) T_\mathrm{eq}}.
\end{equation}
The total excitation $1-P_{\ket{0},\mathrm{ss}}$ is obtained by measurement and postselection~\cite{Riste12b}. Equation~\eqref{eq:gamma01} remains a valid approximation even  for the highest temperatures in Fig.~4, when  the populations of higher excited states become relevant. In this case, the populations $P_{\ket{0},\mathrm{ss}},P_{\ket{1},\mathrm{ss}}$ are estimated from the total excitation, assuming that the populations are thermally distributed~\cite{Riste12b}. 
Error bars for $\Goer, \Gr$ are calculated from the standard deviation of repeated $T_1$ measurements and the fit uncertainty in $\alpha$. 

\section{Acknowledgments}
\begin{acknowledgments}
We thank   G.~Catelani, A.~Endo,  F.~Hassler,  G.~de Lange, J.~M.~Martinis, L.~M.~K.~Vandersypen, P.~J.~de Visser, and the Yale cQED team for discussions. We acknowledge funding from the Dutch Organization for Fundamental Research on Matter (FOM), the Netherlands Organization for Scientific Research (NWO, VIDI scheme), the EU FP7 project SOLID, and the DARPA QuEST program.
\end{acknowledgments}

\end{document}


\title{Supplement to ``Millisecond charge-parity fluctuations and induced decoherence in a superconducting qubit"}
\author{D.~Rist\`e}
\author{C.~C.~Bultink}
\author{M.~J.~Tiggelman}
\author{R.~N.~Schouten}
\affiliation{Kavli Institute of Nanoscience, Delft University of Technology, P.O. Box 5046,
2600 GA Delft, The Netherlands}
\author {K.~W.~Lehnert}
\affiliation{JILA, National Institute of Standards and Technology and Department of Physics, University of Colorado, Boulder, Colorado 80309, USA}
\author{L.~DiCarlo}
\affiliation{Kavli Institute of Nanoscience, Delft University of Technology, P.O. Box 5046,
2600 GA Delft, The Netherlands}
\date{\today}

\maketitle

\section{Experimental setup}
The device and the experimental setup are similar to those described in Refs.~\onlinecite{Riste12,Riste12b}. Here we detail the changes we made since these  earlier reports. To lower the transmon temperature, we replaced the Al cavity with a Cu cavity~\cite{Rigetti12}, improved thermal anchoring to the mixing chamber plate, and added  low-pass filters (K$\&$L Microwave 6L250-8000/T18000-O/O) on the input and output ports of the cavity. As a result, the transmon temperature decreased from $127$ to $55~\mK$, corresponding to a reduction of total steady-state excitation from $\sim16$ to $2\%$, respectively. Because these changes were made simultaneously, we cannot pinpoint the individual contributions to the improved thermalization. 

To perform qubit reset faster, we replaced the ADwin processor with a home-built feedback controller based on a complex programmable logic device (CPLD). 
The CPLD allows a response time, from the end of signal integration to the $\pi$-pulse trigger, of $0.11~\us$. The total loop time, from the start of the measurement pulse to the end of the triggered $\pi$ pulse at the cavity input, is $0.98~\us$, a substantial improvement over the previous $2.62~\us$. 
However, a delay is added to reach $2~\us$ ($\sim10/\kappa$) between measurement and conditioned $\pi$ pulse, ensuring that the cavity is devoid of readout photons. 

\section{Validation of the charge-parity detector}
We here perform several control experiments to validate the use of the qubit as a charge-parity detector. First, the parity-to-qubit state conversion is tested with suboptimal choices of the Ramsey interval $\dt$ (Fig.~S1). As expected from equation (1) in the main text, the white noise level in $S_M$ increases at the expense of the signal contrast as $\dt$ deviates from the optimal choice $1/4\df$. Remarkably, the extracted rate $\Grt$ is approximately constant down to $F~\sim0.4$ (Fig.~S1c). This is consistent with the model of charge parity as a symmetric RTS, with time constant determined solely by QP tunneling. 

\begin{figure*}
\includegraphics[width=2\columnwidth]{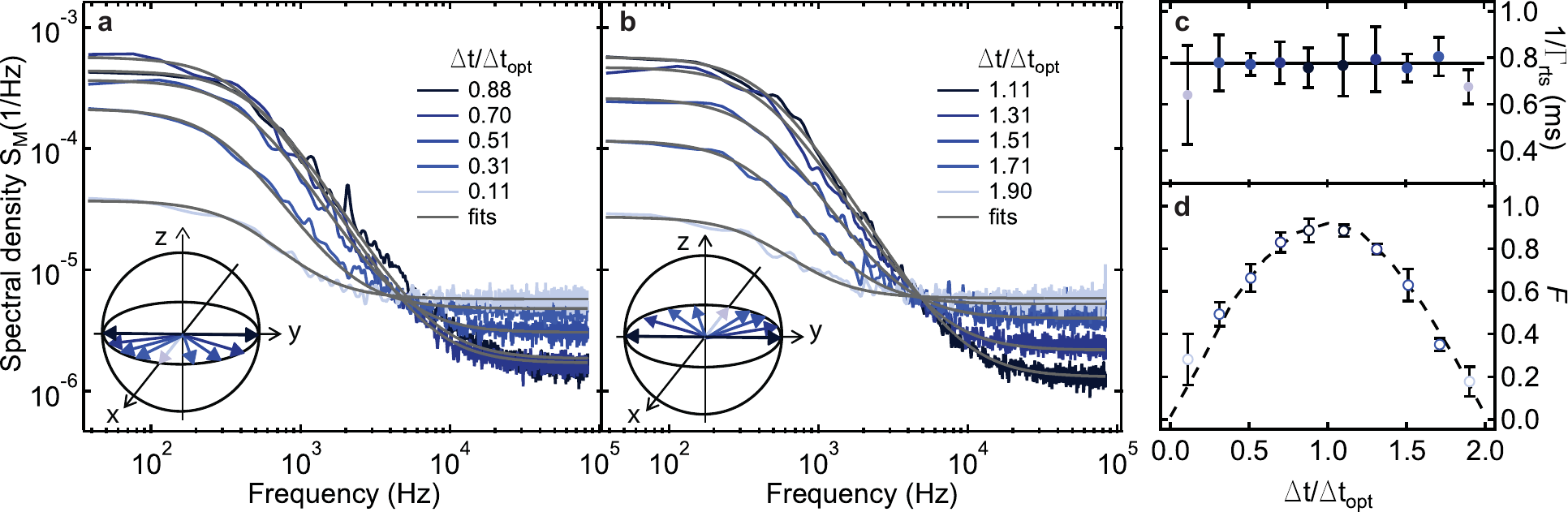}
\caption{{\bf Intentionally imperfect detector}. Power spectral densities similar to those in Fig.~2, but with  Ramsey time $\dt$ varied from the optimum $\dtopt=1/4\df$. The detection fidelity $F$ ({\bf d}) decreases for {\bf a,} $\dt<\dtopt$  and {\bf b,} $\dt>\dtopt$ (b), but the extracted switching time $1/\Grt$ remains approximately the same, as expected  ({\bf c}). Curves in {\bf c} are obtained from a numerical simulation using $1/\Grt=0.78~\ms$.} 
\end{figure*}

In a second test, we replaced the Ramsey-like sequence with a single pulse, with rotation angle $\theta$. Time series of $M$ for $\theta=0$, $\pi$ and $\pi/2$ are shown in Fig.~S2a. The very high occurrence ($\sim99\%$) of $1$ $(-1)$ for $\theta = 0$ $(\pi)$ equals the efficiency of reset, following each measurement $M$. For $\theta = \pi/2$, the qubit is repeatedly prepared in an equal superposition of $\ket{0}$ and $\ket{1}$, and the measurement produces uncorrelated projection noise. The spectra of these control experiments are compared to the QP tunneling measurement in Fig.~S2b, clearly showing that the observed RTS is due to the signal acquired during $\dt$.

\begin{figure*}
\includegraphics[width=2\columnwidth]{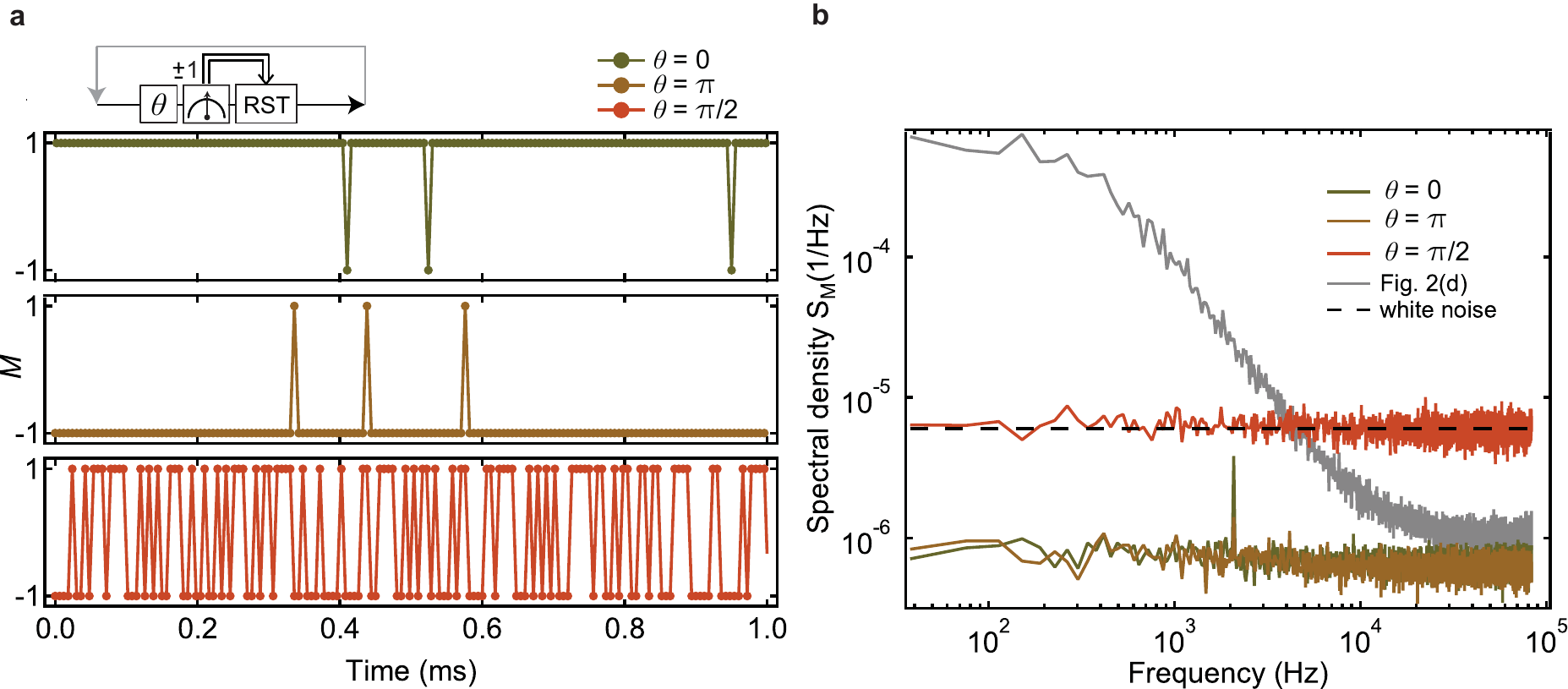}
\caption{{\bf Control experiment replacing the Ramsey sequence in Fig.~2 by a $\theta$-rotation of the qubit}. {\bf a,} $1~\ms$ snapshots of measurements taken at $6~\us$ interval, over $48~\ms$. From top to bottom: $\theta = 0$, $\pi$, and $\pi/2$. For $\theta=0\,(\pi)$, $M=1\,(-1)$, with $\sim1\%$ error, vastly improving on the fast reset demonstrated in Ref.~\onlinecite{Riste12b}. For $\theta=\pi/2$, $M$ is randomized by projection noise. {\bf b,} Power spectral densities of {\bf a}. For $\theta=\pi/2$ the spectrum equals the white noise level (dashed line) at the sampling time of $6~\us$. For $\theta = 0$ or $\pi$, the spectrum is an oder of magnitude lower than the white noise level. The pickup at $2.1~\kHz$ is associated with the pulse tube of the cryogen-free dilution refrigerator, as it disappears when the pulsed-tube compressor is momentarily turned off. The charge-parity spectrum from Fig.~2d is added for comparison.}
\end{figure*}

As a final test of the charge-parity detector, we subject the qubit to an externally generated RTS, similarly to Ref.~\onlinecite{Sun12}. Symmetric RTS sequences  with  switching rate $\Gamma_\pi$ are generated in LabVIEW and sent to an ADwin controller. The ADwin samples the RTS at $9~\us$ interval. When the signal is $+1$, the ADwin triggers an AWG520 (also used for reset~\cite{Riste12b}), which then applies a $\pi$ pulse on the qubit. As a result, the measured qubit state in $M$ is conditioned on the RTS state, mimicking the parity-controlled $\pi$ pulse implemented in Fig.~2. In all cases, the fitted rates $\Gamma_\mathrm{fit}$ match the programmed $\Gamma_\pi$ within $3\%$ (Fig.~S3). 

\section{Slow frequency fluctuations}
The slow drift of the two frequencies evident in Fig.~1c is attributed to rearrangement of background charges on the sapphire substrate, which capacitively couple to the qubit islands. In support of this hypothesis, $\Ttwostar$ follows an upward trend with increasing $\df$ (Fig.~S4a), with maximum ($20-25~\us$) at the charge sweet spot ($\df_\mathrm{max}=880~\kHz$). Histograms of $\df$ (Fig.~S4b) are in good agreement with a sampled sinusoidal function, as expected for charge-modulated qubit frequency. The measured $\Ttwostar$ is consistent in order of magnitude with that estimated from the frequency power spectrum, proportional to $1/f^{1.7}$ over 2 decades. 

\begin{figure}
\includegraphics[width=\columnwidth]{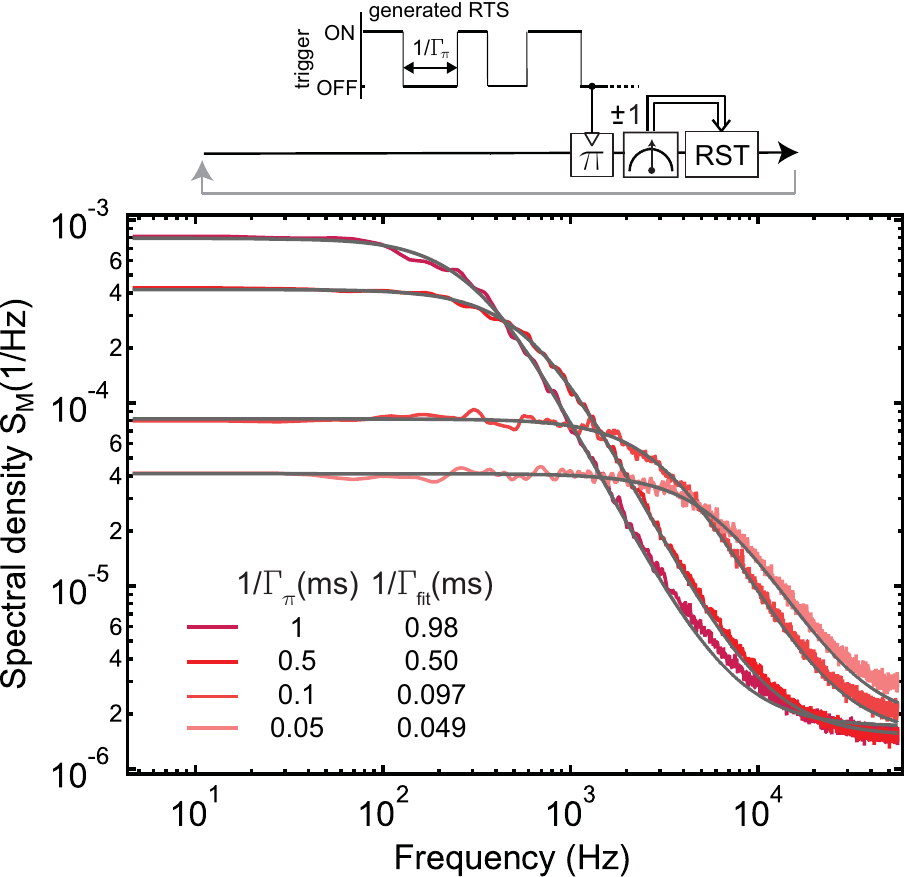}
\caption{{\bf Power spectra of software-generated symmetric RTS, as detected by the qubit}. Programmed sequences of symmetric RTS with switching rate $\Gamma_{\pi}$ are loaded into an ADwin controller that triggers a $\pi$ pulse on the qubit when the level is high (similar to Ref.~\onlinecite{Sun12}). The qubit is then measured and reset to $\ket{0}$ every $9~\us$. This sequence emulates a RTS process similar to QP tunneling. The  detected  and programmed switching rates match to better than $3\%$ over the range $50$ to $1000~\us$.}           
\end{figure}

\begin{figure*}
\includegraphics[width=2\columnwidth]{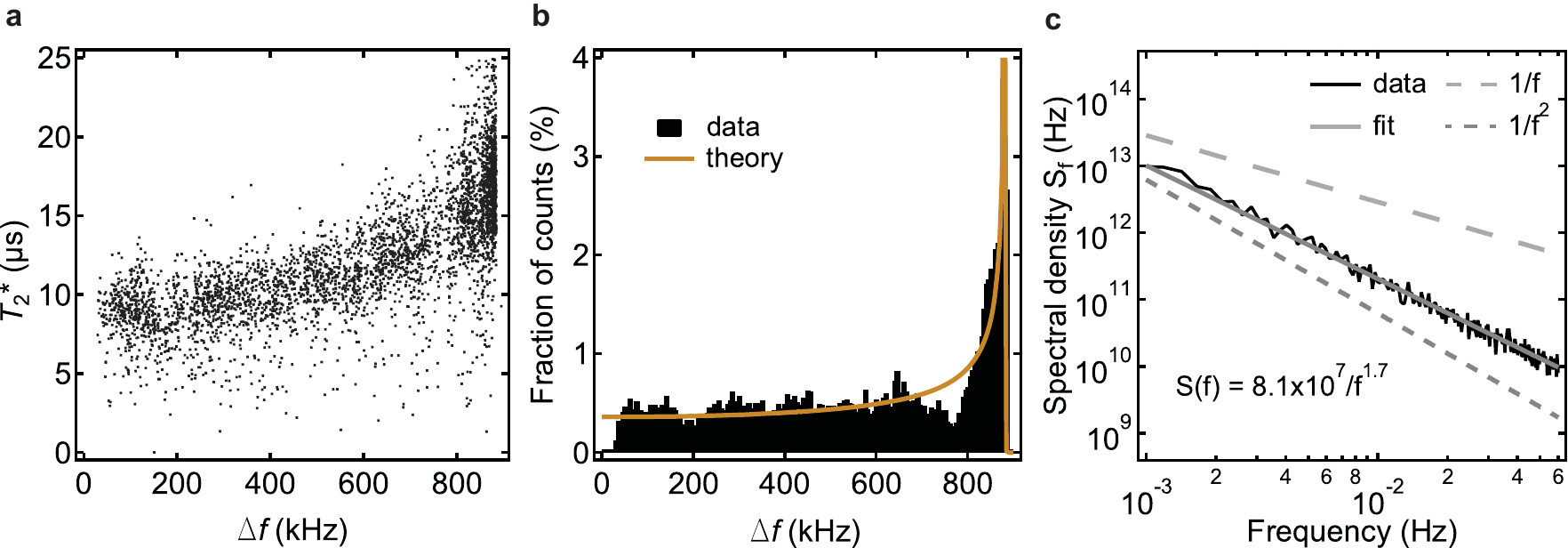}
\caption{{\bf Qubit dephasing and charge noise}. {\bf a,} Scatter plot of  $\Ttwostar$ as a function of $\Delta f$ (see Fig.~1). The upward trend with $\Delta f$ suggests that background charge noise is the main source of qubit dephasing. {\bf b,} Histograms of $\df$, obtained from $30,000$ Ramsey experiments acquired at $\sim4~\s$ interval, with $5~\kHz$ binning. The distribution peaks at the charge sweet spot ($\df_\mathrm{max} = 880~\kHz$), where the qubit transition frequency is least sensitive to background charge fluctuations. {\bf c,} Frequency power spectrum obtained from data in (b). We fit the equation $A/f^\alpha$ to the data for $f>10^{-3}~\Hz$, giving best-fit values $A = 8.1\cdot 10^7$ and $\alpha=1.7$ (solid line). Dashed lines indicate the slopes for $\alpha=1,2$.} 
\end{figure*}

\begin{figure}
\includegraphics[width=\columnwidth]{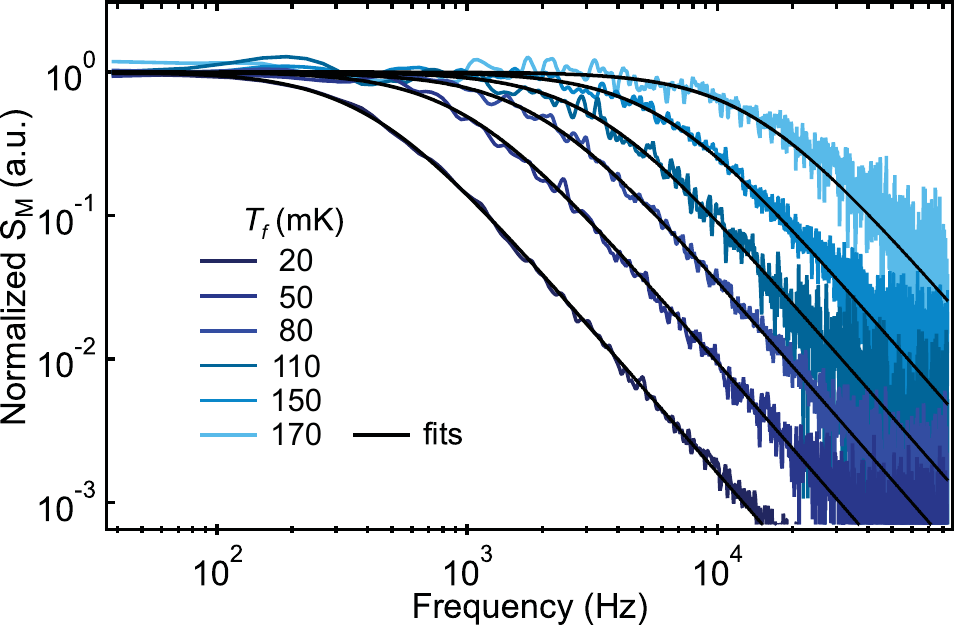}
\caption{{\bf Charge-parity spectra at variable $\Tf$}. Equation~(1) in the main text is fit to each spectrum. To facilitate comparison, data and fits have the white noise offset subtracted and are renormalized. The best-fit rates are plotted in Fig.~4c.} 
\end{figure}